\documentclass[aps,pra,preprint,groupedaddress]{revtex4-1}
\usepackage{braket}
\usepackage[dvips]{graphicx}

\usepackage{amsmath,amssymb}

\newcommand{\etal}{{\em et al.~}}
\newcommand{\pure}[1]{\ket{#1}\!\bra{#1}}

\begin{document}


\title{The Husimi function and a successive measurement of the position and the momentum}


\author{Takuro Shito}
\email[]{sitotaku@quest.is.uec.ac.jp}
\affiliation{Graduate School of Information Systems, University of
Electro-communications, Chofu, Tokyo 182-8585, Japan}


\begin{abstract}
 In this paper, we consider an interpretation of the Husimi function 
as the probability distribution of a successive measurement, 
which is clearly separated into measurements of the position and the momentum. 
We also show this successive measurement can be easily realized in the situation 
close to the experiment, 
and this measurement corresponds to one of Aharanov's weak measurement.
\end{abstract}

\maketitle


\section{introduction}
In this paper, we will discuss properties and structures of the Husimi function 
as the probability distribution of a successive measurement 
of the position and the momentum.

In the literature,
Wigner \etal discussed the distribution of the position and the momentum, 
defined so called the Wigner function \cite{wig1}, 
and considered its properties in 1930s, 
while Husimi \etal discussed the distribution of 
the position and the momentum, 
and defined so called Husimi function \cite{husimi1}.
In the Wigner function, 
there is a problem that the Winger function 
is not limited to a positive function.
Feynman similarly discussed the negative probabilities 
at the spin system \cite{book3}.
Holevo discussed a simultaneous measurement of 
the position and the momentum in 1970s, 
he showed a measurement process of the Husimi function 
by using Naimark extension \cite{book1}.
Recently Aharanov \etal \cite{weakm} proposed 
a weak measurement and a weak value which is similar to the negative probability.

In our study, we discussed the meaning of the marginal distribution of the Husimi function, 
and discovered that the Husimi function can be obtained by Aharanov's weak measurement.
First, we show that the measurement to obtain the Husimi function can be clearly 
separated into two measurements of the position and the momentum.
Secondly, we show that these measurements can be easily realized in the situation close to the experiment.
Finally, we discuss the relation between these measurements and Aharanov's weak measurement.

\section{definition and review of the husimi function}

The Wigner function, the Husimi (Q) function, and the Glauber-Sudarshan (P) function 
are known as distributions including complete information of the quantum state.
In this section, we will define and review the Husimi function 
and other distribution functions \cite{wig1,husimi1,coh1}.
These distribution functions are defined by the following 
family of characteristic functions parameterized by a parameter $s$:
\begin{equation}
\tilde{w}(u,v,s)=\textmd{Tr}\hat{\rho}e^{-iu\hat{x}-iv\hat{p}}e^{\frac{s}{4}(u^2+v^2)},
\end{equation}
where $\hat{\rho}$ is the density operator of a state satisfying
$\hat{\rho}\geq 0$ and $\textmd{Tr}\hat{\rho}=1$.
Using these characteristic functions, we define
\begin{eqnarray}
W_{\hat{\rho}}(x,p)&=&\frac{1}{(2\pi)^2}\iint dudv
\tilde{w}(u,v,0)e^{iux+ivp},\\
Q_{\hat{\rho}}(x,p)&=&\frac{1}{(2\pi)^2}\iint dudv
\tilde{w}(u,v,-1)e^{iux+ivp},\label{def_husi2}\\
P_{\hat{\rho}}(x,p)&=&\frac{1}{(2\pi)^2}\iint dudv \tilde{w}(u,v,+1)e^{iux+ivp},
\end{eqnarray}
where $W_{\hat{\rho}}(x,p)$ is the Wigner function ($s=0$),
$Q_{\hat{\rho}}(x,p)$ is the Husimi function ($s=-1$),
and $P_{\hat{\rho}}(x,p)$  is the Glauber-Sudarshan function ($s=1$).
Here and hereafter the range of the integral is 
from $-\infty$ to $\infty$ and is omitted.

While these functions are defined for density operators,
we can also define them with respect to any linear operators by linearity.
Moreover the trace of the product of any two linear operators 
can be written by Wigner functions.
\begin{equation}
\textmd{Tr}\hat{A}\hat{B}=\iint dxdp W_{\hat{A}}(x,p) W_{\hat{B}}(x,p).
\end{equation}
Using this fact, the expectation of any observable $\hat{A}$ 
for a state $\hat{\rho}$ is written by Wigner functions.
\begin{equation}
\braket{\hat{A}}_{\hat{\rho}}=\textmd{Tr}\hat{\rho}\hat{A}=\iint dxdp
W_{\hat{\rho}}(x,p) W_{\hat{A}}(x,p).
\end{equation}
Similarly, using the Husimi function and the Glauber-Sudarshan function,
we can also write the expectation of an observable $\hat{A}$ as follows:
\begin{eqnarray}
\braket{\hat{A}}_{\hat{\rho}}=\textmd{Tr}\hat{\rho}\hat{A}
&=&\iint dxdp Q_{\hat{\rho}}(x,p) P_{\hat{A}}(x,p)\label{ex_by_gs}\\
&=&\iint dxdp P_{\hat{\rho}}(x,p) Q_{\hat{A}}(x,p).\label{ex_by_husimi}
\end{eqnarray}

We can derive a simple formula of the Wigner function and the Husimi function 
for a state $\hat{\rho}$ respectively,
\begin{eqnarray}
W_{\hat{\rho}}(x,p)&=&\frac{1}{2\pi}\int dy
e^{ipy}\bra{x-y/2}\hat{\rho}\ket{x+y/2},\\
Q_{\hat{\rho}}(x,p)&=&\frac{1}{2\pi}\bra{x,p}\hat{\rho}\ket{x,p},
\label{husimi}
\end{eqnarray}
where $\ket{x,p}$ is a ``coherent state,'' i.e., 
an eigenvector of the annihilation operator $\hat{a}$.
It can be seen from \eqref{husimi} that the Husimi function is always positive,
while the Wigner function and the Glauber-Sudarshan function 
sometimes take negative values.
The marginal distribution of the position $x$ 
for the Husimi function and the Wigner function is, respectively, written by
\begin{eqnarray}
\int dp W_{\hat{\rho}}(x,p)&=&\bra{x}\hat{\rho}\ket{x},\\
\int dp Q_{\hat{\rho}}(x,p)&=&\frac{1}{\sqrt{\pi}}\int dx^\prime
e^{-(x-x^\prime)^2}\bra{x^\prime}\hat{\rho}\ket{x^\prime}\label{mhd1}.
\end{eqnarray}
In the subsequent discussion, we consider properties of the Husimi function.
The equation \eqref{husimi} is frequently used as a definition of the Husimi function.
In our discussion we follow this definition,
because this definition can include a freedom of the squeezed parameter $\Delta$ as follows.
A squeezed coherent state is expressed by the vacuum state $\ket{0}$,
 which is a minimum uncertainty state, i.e., the state satisfying $\Delta x\Delta p=1/4$.
Let $\hat{S}(\Delta)$ be the squeezed operator and $\hat{D}(x,p)$ be the shift operator. 
Then a squeezed coherent state is expressed by
\begin{eqnarray}
\ket{x,p;\Delta}&=&\hat{D}(x,p)\hat{S}(\Delta)\ket{0}\nonumber\\
&=&\frac{1}{(\Delta\pi)^{1/4}}\int dx^\prime
e^{-\frac{1}{2\Delta}(x^\prime-x)^2+ix^\prime p}\ket{x^\prime}.
\label{s-coherent}
\end{eqnarray}
Now the Husimi function based on the squeezed coherent state is defined by
\begin{equation}
Q(x,p;\Delta)=\frac{1}{2\pi}\bra{x,p;\Delta}\hat{\rho}\ket{x,p;\Delta}.
\label{def:husimi}
\end{equation}
When the squeezed parameter $\Delta = 1$, 
this definition corresponds to the definition by the characteristic function \eqref{def_husi2}.
In the following, we omit the squeezed parameter $\Delta$ in $\ket{x,p;\Delta}$ 
for simplicity.

\section{husimi distribution function and post selection}
In this section we study the detail of a physical interpretation of the Husimi function.
Considering a pure state $\hat{\rho}=\pure{\psi}$ 
in the definition of the Husimi function \eqref{def:husimi},
we have $Q(x,p;\Delta)=\frac{1}{2\pi}|\braket{x,p;\Delta|\psi}|^2$.
The term $\braket{x,p;\Delta|\psi}$ can be calculated as follows:
\begin{eqnarray}
\braket{x,p;\Delta|\psi}&=&\frac{1}{(\Delta\pi)^{1/4}} \int dx^\prime
e^{-\frac{1}{2\Delta}(x-x^\prime)^2-ix^\prime
p}\braket{x^\prime|\psi}
\label{eq15}\\
&=&\sqrt{2\pi}\bra{p}\frac{1}{(\Delta\pi)^{1/4}}
e^{-\frac{1}{2\Delta}(x-\hat{x})^2}\ket{\psi},
\label{eq16}
\end{eqnarray}
where we used the spectral decomposition of $\hat{x}$:
\begin{equation}
\hat{x}=\int dx^\prime x^\prime \pure{x^\prime}.
\end{equation}
Now let us define an operator
$\hat{M}(x)=\frac{1}{(\Delta\pi)^{1/4}}e^{-\frac{1}{2\Delta}(x-\hat{x})^2}$,
then \eqref{eq16} is rewritten as
\begin{equation}
\braket{x,p|\psi}=\sqrt{2\pi}\bra{p}\hat{M}(x)\ket{\psi}.\label{sep_husi}
\end{equation}
In the following, we will interpret the square of the absolute value of the above equation 
as the probability of a successive measurement of the position and the momentum.
We will also discuss a physical realization of the successive measurement in the next section.
The operator $\hat{M}(x)$ is a measurement operator, 
which fulfill flowing properties;
\begin{equation}
\int dx\hat{M}^2(x)=\hat{I},
\end{equation}
and the operator $\hat{M}(x)$ mean the reduction of the wave packet, thus the state after 
the measurement is;
\begin{equation}
\ket{\psi^\prime}\propto\hat{M}(x)\ket{\psi}=\frac{1}{(\Delta\pi)^{1/4}}\int dx^\prime
\psi(x^\prime)e^{-\frac{1}{2\Delta}(x-x^\prime)^2}\ket{x^\prime},
\end{equation}
where $\psi(x)=\braket{x|\psi}$.
This operator $\hat{M}(x)$ have other properties described below.
The coherent state $\ket{x,p}$ is generated by operating $\hat{M}(x)$ 
to some eigenstates of the momentum $\ket{p}$.
\begin{equation}
\sqrt{2\pi}\hat{M}(x)\ket{p}=\ket{x,p}
\end{equation}
When we measure the eigenstate of the momentum $p$ with $\hat{M}(x)$, 
the state will become the coherent state after the reduction of the wave packet.
The expectation of the squared operator $\hat{M}(x)$ is 
the marginal distribution of the Husimi function.
\begin{eqnarray}
\bra{\psi}\hat{M}^2(x)\ket{\psi}&=&\bra{\psi}\frac{1}{\sqrt{\Delta\pi}}
e^{-\frac{1}{\Delta}(x-\hat{x})^2}\ket{\psi}\nonumber\\
&=& \frac{1}{\sqrt{\Delta\pi}}\int dx^\prime
e^{-\frac{1}{\Delta}(x-x^\prime)^2}|\braket{x^\prime|\psi}|^2\\
&=&\int dp Q(x,p)\label{marginalq}
\end{eqnarray}
However, this distribution does not have complete information of the state $\ket{\psi}$.
The operator $\hat{M}(x)$ is written by a coherent state $\ket{x,p}$ as follows:
\begin{equation}
\hat{M}(x)=\bigg\{\int dp
\ket{x,p}\bra{x,p}\bigg\}^{\frac{1}{2}}.
\label{M(x)}
\end{equation}
After we obtain the measurement result $x$ of $\hat{M}(x)$, we have one to one correspondence 
between the the state after the measurement $\ket{\psi^\prime}$ and the initial state $ket{\psi}$.
In this sense, the state after measurement $\ket{\psi^\prime}$ does not lose 
any information of the initial state $\ket{\psi}$.

\section{Physical realization of the measurement
of the position and post selection of the momentum}

In this section, we consider a physical realization of the
measurement $\hat{M}(x)$ in \eqref{M(x)}.  In the following, we show
that the Husimi function is obtained as a probability distribution of a
successive measurement, i.e., a weak measurement of the position
followed by the projective measurement of the momentum.
Note that these measurements of the position and the momentum
are quite different in the sharpness,
while the Husimi function as the probability distribution
has symmetry in the position and the momentum.
It is known \cite{book1} that
a simultaneous measurement of the position and the momentum is possible
on an enlarged Hilbert space by Naimark extension.
We can easily see that 
the measurement devised by Holevo, is corresponding to one of measurements 
which obtain the Husimi function.

We observe some properties of the position measurement \eqref{M(x)}
related to the squeezed parameter $\Delta$ in \eqref{s-coherent}.
First, when $\Delta\rightarrow0$, i.e., the position measurement is sharp,
following probability distribution $p(x)$ is obtained for any state $\ket{\psi}$;
\begin{eqnarray}
p(x)&=&\lim_{\Delta\rightarrow0}\bra{\psi}\hat{M}^2(x)\ket{\psi}\nonumber\\
&=&\lim_{\Delta\rightarrow0}\int dx^\prime \frac{1}{\sqrt{\Delta\pi}}
e^{-\frac{1}{\Delta}(x-x^\prime)^2}|\braket{x^\prime|\psi}|^2\nonumber\\
&=&\int dx^\prime \delta(x-x^\prime)|\braket{x^\prime|\psi}|^2\nonumber\\
&=&|\braket{x|\psi}|^2.
\end{eqnarray}
After this measurement, the initial state $\ket{\psi}$ will be 
the eigenstate of the position $\ket{x}$, 
which corresponds with obtained value $x$. 

Secondly,
when $\Delta\rightarrow\infty$, that is
the unsharp limit of the position measurement,
\eqref{M(x)} corresponds with the identity operator
for any $x$
\begin{eqnarray}
\lim_{\Delta\rightarrow\infty}\hat{M}(x)
&=&\lim_{\Delta\rightarrow\infty}\frac{1}{(\Delta\pi)^{1/4}}e^{-\frac{1}{2\Delta}(x-\hat{x})^2}\nonumber\\
&\propto&\lim_{\Delta\rightarrow\infty}\int dx^\prime e^{-\frac{1}{2\Delta}(x-x^\prime)^2}\ket{x^\prime}\bra{x^\prime}\nonumber\\
&\simeq&\int dx^\prime\ket{x^\prime}\bra{x^\prime}\nonumber\\
&=&\hat{I}
\label{unsharp-M(x)}
\end{eqnarray}
Thus we can not obtain any information of the system $\ket{\psi}$ 
by the unsharp measurement $\lim_{\Delta\rightarrow\infty}\hat{M}(x)$.
If we measure the momentum
after the position measurement $\lim_{\Delta\rightarrow\infty}\hat{M}(x)$,
the probability distribution is obtained as follows:
\begin{eqnarray}
|\bra{p}\lim_{\Delta\rightarrow\infty}\hat{M}(x)\ket{\psi}|^2&\propto&|\bra{p}\hat{I}\ket{\psi}|^2=|\braket{p|\psi}|^2.
\end{eqnarray}

We show that the measurement $\hat{M}(x)$ can be realized by a von Neumann
measurement as follows.
We consider a composite system of the
measurement device
with squeezed vacuum state $\ket{0;\Delta}$
and the measured system with an initial state $\ket{\psi}$;
\begin{equation}
\ket{0;\Delta}\ket{\psi}=\int dx^\prime e^{-\frac{1}{2\Delta}x^{\prime2}}\ket{x^\prime}\ket{\psi},\label{init}
\end{equation}
and the time evolution by the following interaction Hamiltonian;
\begin{equation}
\hat{H}_{inter}=-\delta (t-t^\prime)\hat{x}_\psi\hat{p}_{md},
\label{Hamiltonian}
\end{equation}
where $\hat{x}_\psi$ operate on the measured system, $\hat{p}_{md}$ operate
on the system of the measurement device
and $\hat{x}_\psi\hat{p}_{md}$ means $\hat{x}_\psi\otimes\hat{p}_{md}$.
The position of the measurement device is shifted
by this interaction Hamiltonian at $t=t^\prime$
in the proportion to the position of the measured system.

The unitary operator of the time evolution
between $t=0$ and $t=t^\prime$ is written as
\begin{equation}
\hat{U}=\exp \bigg\{ \int^{t^\prime}_0 dt \hat{H}_{inter} \bigg\}=\exp\{-i\hat{x}_\psi\hat{p}_{md}\}.
\label{unitary}
\end{equation}
After the interaction \eqref{unitary}, the composite system \eqref{init} will be written as
\begin{equation}
\int dx^\prime e^{-\frac{1}{2}(x^\prime-\hat{x}_\psi)^2}\ket{x^\prime}\ket{\psi}.
\end{equation}
We measure the position of the measurement device
and consecutively measure the momentum of the measured system,
then the probability of the position and the momentum is given by
\begin{align}
\left|
\bra{p}\bra{x}\int dx^\prime
e^{-\frac{1}{2}(x^\prime-\hat{x}_\psi)^2}\ket{x^\prime}\ket{\psi}
\right|^2
=\left|
\int dx^{\prime\prime}e^{-\frac{1}{2\Delta}
(x-x^{\prime\prime})^2-ix^{\prime\prime}p}\braket{x^{\prime\prime}|\psi}
\right|^2,
\label{md_wf}
\end{align}
which corresponds to the Husimi function of the measured system \eqref{eq15}.

In the following we show that this measurement and 
Aharanov's weak measurement \cite{weakm} are related.
Let us introduce a Hamiltonian corresponding to \eqref{Hamiltonian} by
\begin{equation}
\hat{H}_{inter}=-g\delta (t-t^\prime)\hat{x}_\psi \hat{p}_{md}.
\label{Hamiltonian2}
\end{equation}
Note that the above Hamiltonian
includes the parameter $g$ which means the weakness of the interaction.
We also consider the composite system
corresponding to \eqref{init}, without parameter $\Delta$:
\begin{equation}
\ket{0}\ket{\psi}=\int dx^\prime e^{-\frac{1}{2}x^{\prime2}}\ket{x^\prime}\ket{\psi}.\label{init2}
\end{equation}
After the interaction by the Hamiltonian \eqref{Hamiltonian2},
the composite system \eqref{init2} will be
\begin{equation}
\int dx^{\prime\prime} e^{-\frac{1}{2}(x-gx^{\prime\prime})^2-ix^{\prime\prime}p}\braket{x^{\prime\prime}|\psi}.
\end{equation}
Then we measure the position of the measurement device
and the momentum of the measured system.
\begin{eqnarray}
\int dx^{\prime\prime} e^{-\frac{1}{2}(x-gx^{\prime\prime})^2-ix^{\prime\prime}p}\braket{x^{\prime\prime}|\psi}
=
\int dx^{\prime\prime} e^{-\frac{g^2}{2}(x/g-x^{\prime\prime})^2-ix^{\prime\prime}p}\braket{x^{\prime\prime}|\psi}
\end{eqnarray}
Now we assume that $\Delta=1/g^2$ and $x/g$ is $\bar{x}$, then we obtain
\begin{equation}
\int dx^{\prime\prime} e^{-\frac{1}{2\Delta}(\bar{x}-x^{\prime\prime})^2-ix^{\prime\prime}p}\braket{x^{\prime\prime}|\psi}.
\end{equation}
This probability distribution corresponds with \eqref{md_wf}.
The Husimi function defined in \eqref{def:husimi} includes the controllable parameter $\Delta$, 
which means the sharpness of the position measurement,
while Aharanov \etal supposed that the interaction between 
the measured system and the measurement device is so weak.

The momentum measurement in our measurement is the projective measurement,
and is clearly separated from the position measurement.
Thus we can easily see that the conditional
probability distribution of the position, when the measured value of the momentum is
$p$, is given by
\begin{equation}
Q(x|p)=\frac{Q(x,p)}{\bra{\psi}(\ket{p}\bra{p})\ket{\psi}},
\label{cp_by_husimi}
\end{equation}
where $Q(x,p)$ is the Husimi function.

Holevo \cite{book1} already discussed a measurement 
to obtain the Husimi function by Naimark extension. 
However, our measurement is different form Holevo's measurement 
in the point that Holevo's measurement is simultaneous measurement 
in the extended Hilbert space by Naimark extension.
Additionally, in our measurements of the position and the momentum, 
each measurements are clearly separated.
On the other hand Holevo's measurements 
are not separated, because   
the measurement of the relative position and the center momentum 
across the extend Hilbert space are considered.
Using our measurement, we can give comprehensible physical meaning to equation \eqref{cp_by_husimi}.

\section{conclusion}
We proposed a new interpretation of the Husimi function 
including post-selection studying 
the marginal distribution of the Husimi function and Aharanov's weak measurement.
We also showed that the probability distribution of 
our successive measurement of the position and the momentum 
corresponds to the Husimi function. 
This measurement process is different from Holevo's measurement 
in the point that his measurement is a simultaneous measurement 
across the extended  Hilbert space, while our measurement is a separated successive measurement.
The important point of our study is that 
our measurement can be realized by the reduction of the wave packet of the measured system 
without considering the measurement device system. In the other words, our each measurements, the position and the momentum, are separated.
By using this fact, 
we can easily calculate the marginal distribution or 
the conditional probability form the Husimi function like \eqref{marginalq} 
and \eqref{cp_by_husimi}.
Although our measurement is the simple combination of the weak measurement of the position 
and the projective measurement of the momentum, the probability distribution
automatically corresponds to the Husimi function by the reduction of the wave packet.
We have also shown that our measurement process to obtain the Husimi function 
is the weak measurement of the position 
with post-selection of the momentum.

\section*{acknowledgement}
The author thanks Dr.Shimada and Dr.Ogawa for helpful and discussion.

\end{document}